# Recurrence plots of exchange rates of currencies

Amelia Carolina Sparavigna

Department of Applied Science and Technology, Politecnico di Torino, Torino, Italy

**Abstract** Used to investigate the presence of distinctive recurrent behaviours in natural processes, the recurrence plots can be applied to the analysis of economic data, and, in particular, to the characterization of exchange rates of currencies too. In this paper, we will show that these plots are able to characterize the periods of oscillation and random walk of currencies and enhance their reply to news and events, by means of texture transitions. The examples of recurrence plots given here are obtained from time series of exchange rates of Euro.

1. **Introduction**

In finance, an exchange rate between two currencies is the rate at which one currency is exchanged for another. The problem of forecasting these rates is highly debated and it is studied with several methods. However, from the examination of the scientific literature, it appears that different forecasting horizons lead to different forecasting performances [1]. Being the exchange rates considered as quantities that are responding immediately to changes in expectations, their behavior is viewed similar to that of stock prices, which are responding strongly to news, in particular to unexpected economic and political events [2,3]. Like stock prices, the exchange rates are hard to forecast, but it is possible to give some trends of them [4].

Using the time series of exchange rates, trends can be calculated using linear or quadratic models and autoregressive models [4]. Let us remember that time series are sequences of data measured typically at successive points, which are spaced at uniform time intervals. These series can be represented in a Cartesian plane, plotting them over time. The horizontal axis is used to plot time, the vertical axis is used to plot the observed values. Besides the Cartesian plots, it is possible to display the time series using recurrence plots. These plots are showing the recurrence of values of the observed quantity, that is, they are showing when values are arbitrarily close again after some time of divergence. These plots were proposed in 1987, by Eckmann, Kamphorst and Ruelle to investigate several natural processes [5,6]. In the case we are using them for currencies, we can see the recurrence at a different time $j$ of the value of the currency observed at time $i$.

A recurrence plot can appear as a two-dimensional squared black and white image, with black dots marking recurrence, the resulting image depending on the threshold distance used to determine the distances between values. However, several tools are freely available to have the recurrence plots given with colours and different layouts to represent distances between values.

Recently, we have discussed the use of recurrence plots in a paper where we studied and compared the oscillations of the levels of some lakes in Africa [7]. As shown in the given reference, the recurrence plots help displaying differences and similarities of the local climatic behaviour of drainage basins. For instance, we found the plot of level oscillations of Lake Kainji similar to the plot of a Rossler attractor [7]. Here, we will show some examples of recurrence plots applied to exchange rates of currencies. The data used for plots are real time series of exchange rates of Euro. We will see the plots able to characterize the periods of oscillation of the rates about fixed values, and those of random walk. Moreover, we will see how plots are able to enhance the observation of changes in expectation and interest of the markets. In particular, we will see that texture transitions, that is, visual discontinuities in the plot, are marking how strongly are the currencies responding to news and events. Before showing the recurrence plots, let us discuss some facts about the exchange rates.

2. **Exchange rates determination**

As explained by the Sauder School of Business, University of British Columbia, [4], and in several books [8-11], long term movements of exchange rates are driven by fundamental forces. On the other hand, short term movements are driven by news and events, and their horizons are days or weeks long. Fundamental economic forces, such as purchasing power parity (PPP) and balance-of-payment disequilibrium, usually take much longer, often several years, to have an effect on exchange rates. The purchasing power parity is considered the leading force by those long-time theories which state that the exchange rates between currencies are in equilibrium when their purchasing power is the same in each of the two countries.

Besides the theories suitable for long-time scenarios, the business community need models for forecasting short-term trends of exchange rates. Since the related expectations are created by market sentiments, that is, the overall attitude of investors toward a particular financial market, trends can be extrapolated. However, intrinsic difficulties exist because the market sentiments, which are not always based on fundamentals, can change when the outlook on a country's economic prospects changes.

Models giving short-term trends of exchange rates exist: however, trends are not forecasts [4], because they can be broken at any time due to the arrival of new information. Trends can be calculated using linear or quadratic models and autoregressive models, such as the stepwise autoregressive method [4]. Let us suppose we have time series data for the past several months and that we want the trend for the next months. We can apply a statistical analysis based on a linear or quadratic model, in order to capture long-term behavior, then we can use an autoregressive component, to have information on short-term behavior.

In fact, the long-term movements in time series data can be easily represented by trend lines, which are determined using statistical techniques such as the linear regression. These lines tell whether the data set increased or decreased over the considered period of time. Usually, they are straight lines, although higher degree polynomials can be used too, depending on the supposed degrees of curvature. Once the line is obtained, we can tell the trend. To have a prediction of the future values, autoregressive models are used.

3. **Autoregressive models**

Autoregressive (AR) models are representing some time-varying processes where the output variable depends linearly on its own previous values. An AR model of the first order can be defined as:

$$X_i = c + \rho X_{i-1} + e_i \quad (1)$$

Here we have a constant $c$, a parameter $\rho$ and a white noise $e_i$. Index $i$ is representing the time interval or time period. In (1), we see the AR(1) model where the value $X_i$ is depending just on the previous value $X_{i-1}$ in the time series. That is, in the case of currencies, the value of today depends on the value of yesterday. However, in general we can have a AR(p) model:

$$X_i = c + \sum_{j=1}^{p} \rho_j X_{i-j} + e_i \quad (2)$$

In this case, the value of today can depend on the values of several previous days. The number of parameters is *p*. Of course, the simplest cases are those having one or two parameters, that is, AR(1) or AR(2).

Let us note that in an AR process, a one-time shock affects the values of *X* infinitely far into the future. In the AR(1) model (1) for instance, a non-zero value at say $i=1$ of $e_i$ affects $X_1$ by the amount $e_1$. Then, at $i=2$, due to (1), this shock affects $X_2$ by the amount $\rho e_1$. Continuing, $X_3$ is affected by the amount $\rho^2 e_1$ and so on. This process in fact never ends. However, in the case the process is stationary, the effect diminishes toward zero.

Once the parameters of the autoregressive model have been estimated in some manner, the model (1) or (2) can be used to forecast into the future, determining *X* the first time for which the datum is not yet available, using the previous known values. The output of the autoregressive equation is the forecast for the first unobserved period. It is possible to continue, and use (1) or (2) for the further next periods for which data are not yet available. We are then making further forecasts. In this case, the value of *X* is predicted using the previous forecasts.

The obtained predictions are affected by some uncertainties. First, we have to consider whether the autoregressive model is the correct model or not. Then, we have the uncertainty about the accuracy of the number *p* of parameters to use and the uncertainty about the true values of the autoregressive coefficients. Moreover, we have to decide the error term *e* for the period being predicted.

With the moving average (MA) models, AR models are the most used models. In 1976, George Box and G.M. Jenkins proposed a methodology by combining AR and MA models to produce the ARMA (autoregressive moving average) model, fundamental for studying the stationary time series. ARIMA (autoregressive integrated moving average models) on the other hand are used to describe non-stationary time series [12]. In econometrics, autoregressive conditional heteroskedasticity (ARCH) models are used too. The term "heteroskedastic" means "with differing variance" and comes from the Greek "hetero" meaning "different" and "skedasis" meaning "dispersion". The ARCH models assume the variance of the current error term being a function of the actual sizes of the previous error terms. ARCH models are employed commonly in modelling financial time series that exhibit time-varying volatility clustering, that is, periods of swings followed by periods of relative calm [13].

### 4. Use of recurrence analysis

Autoregressive models are popular for modelling the dynamics of real exchange rates, in particular to examine the validity of the purchasing power parity theory [14]. As told in [15], the fulfilment of the purchasing power parity is imposing equilibrium restrictions to the evolution of nominal exchange rates, and therefore, that the real exchange rates must follow a stationary process. However, the most popular linear models usually yield the finding that the real exchange rates are non-stationary [14]. To solve this difficulty, various nonlinear models have been proposed: among them the band-threshold autoregressive (TAR) models, where the real exchange rates are random walks inside bands of inaction, the exponential smooth transition autoregressive (ESTAR) models and the GARCH models, that are generalized autoregressive conditional heteroskedasticity models [14].

It was the investigation of the true nature of underlying processes in exchange rates that inspired a huge number of studies concerning nonlinear dynamics in economic data. As stressed in [15], several researchers considered the possibility that the underlying processes could be nonlinear stochastic or chaotic. Since it is very important to know if deterministic structures, chaotic or not, exist in the exchange rates, methods able to investigate such situation are necessary; some of them are based on recurrence analysis. In [15] then, the use of recurrence plots and analysis had been considered and discussed. The aim of authors was that of checking the possibility of chaos. In fact, a chaotic dynamics could explain why purchasing power parity does not hold for some exchange rates when applying traditional tests [15].

In [16], the programs available to apply the recurrence plot methodology have been discussed, in particular the Visual Recurrence Analysis (VRA). We have proposed its use in [7]: as previously told, we can have two-dimensional squared matrix with black and white dots or coloured images using VRA, a software package for analysis, qualitative and quantitative assessment, and for nonparametric prediction of nonlinear and chaotic time series [17,18].

## 5. Recurrence plots of Euro

Let us consider the time series of some exchange rates of Euro. The real data are provided by [19,20]. The time series we consider for plotting contain the daily values spanning the period from July 1, 2010 to July 1, 2014. Of course, several currencies could be analyzed: here we use US Dollar, GB Pound and Japanese Yen.

Let us consider the date $i$ and the corresponding value of the exchange rate $X_i$. The procedure to obtain a recurrence plot $RP$ is given by:

$$RP_{i,j} = \Theta(T - |X_i - X_j|) \quad (3)$$

$X_j$ is the value on date $j$. $T$ is a threshold. The $\Theta$-function is 0 if the absolute value of the difference between $X_i$ and $X_j$ is greater than the given threshold $T$, and 1 if the difference is lower $T$. The threshold is defined as:

$$\overline{X} = \frac{1}{N}\left(\sum_{i=1}^{N} X_i\right); \quad T = \frac{1}{N}\left(\sum_{i=1}^{N}(X_i - \overline{X})^2\right)^{1/2} \quad (4)$$

$N$ is the number of data. The threshold $T$ is considered as the average of the distances of each value from the average value $\overline{X}$.

The recurrence plot is displayed in an image where 1 is represented by a black dot and 0 by a white dot. In the Figure 1 we see the recurrence plots for the exchange rates between Euro and US Dollar and between Euro and GB Pound.

These recurrence plots are typical of autoregressive systems. Using the plot defined with (3) and (4), we see values having distances lower than the threshold. In fact, in the Figure 1 we can see that there are periods of oscillation in the recurrence plots, represented by checkerboard textures. It means that in these periods, the value is oscillating about one or two fixed points [5]. We have also the texture with circles typical of random walks [21]. After having observed oscillations and random walks in the recurrence plot, we could investigate the economic and politic facts of the corresponding periods of time, to evaluate the relevance of them on the exchange rates.

In the Figure 2 we have the two recurrence plots of Figure 1 shown together. In blue we see the exchange rate Euro - US Dollar and in red the exchange rate Euro - GB Pound. The image shows that the periods of time when the two rates oscillate are often coincident, such as the random walks.

Instead of (3) and (4), we can use VRA tool where all the distances are represented by colours. In the Figure 3, we see the VRA recurrence plots of the same time series, where colours are representing an Euclidean distance.

### 6. The case of the Japanese Yen

Let us consider the recurrence plot of exchange rates between Euro and Japanese Yen. If we look at the upper part of Figure 4, where we can see a recurrence plot based on Eqs.3 and 4, the behavior of the exchange rate is more or less the same, with oscillations and random walks. In fact, year 2014 is a period of oscillation for the currencies we are examining.

Let us plot the exchange rate with VRA: in the lower part of Figure 4 we can see a clear texture transition on the date of April 4, 2013 (this transition cannot be observed using the previous method). With texture transition we consider a visual discontinuity, such as a misalignment with adjacent geometries [22]. In fact, texture transitions had been detected in liquid crystals too [23,24].

We can ask ourselves the reason for this texture transition. As told in [25], on April 4, 2013, the Governor Haruhiko Kuroda "committed the BOJ (Bank of Japan) to open-ended asset buying and said the monetary base would nearly double to 270 trillion yen ($2.9 trillion) by the end of 2014, a dose of shock therapy officials hope will end two decades of stagnation". And financial markets liked this. "The yen fell more than 3 percent against the dollar and 4 percent against the euro, while the 10-year government bond yield hit a record low" [25]. This is what we see in the texture transition in the VRA recurrence plot in Figure 4.

### 7. Conclusion

The paper shows the recurrence plots obtained from time series of exchange rates between some currencies and Euro. The plots have been prepared in two different manners. One gives black and white images, the second is using VRA tool producing coloured images. In both cases, we can clearly see the periods of oscillations, represented by checkerboard textures and the circles of random walks. Using VRA tool we can see the presence of texture transitions, that is, the presence of visual discontinuities in the pattern or texture of the recurrence plot. Therefore, the VRA plots are able to enhance the observation of rapid changes in expectations and interest of the markets. An example is the recurrence plot of exchange rate EUR/JPY, where we can see a texture transition on April 4, 2013, when markets boosted by the announcements of the Bank of Japan. As shown by the proposed examples, the recurrence plots can be used to visualize periods of oscillation and random walk and to point out how strongly are the currencies responding to news.

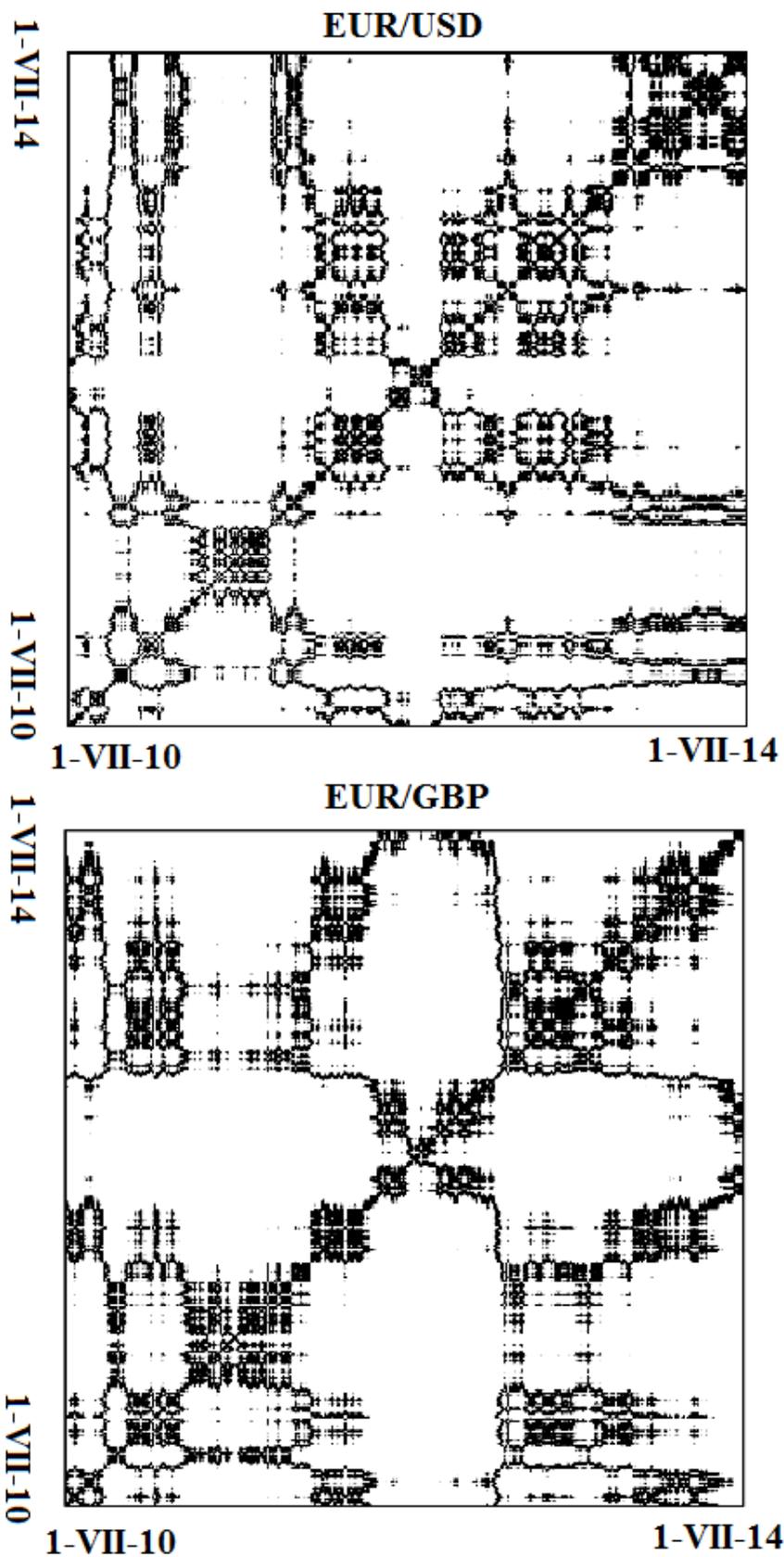

**Figure 1** - Daily exchange rate Euro - US Dollar and Euro - GB Pound from July 1, 2010 to July 1, 2014. We can see several periods of oscillation, represented by checkerboard textures, and periods of random walks with a texture having typical circles.

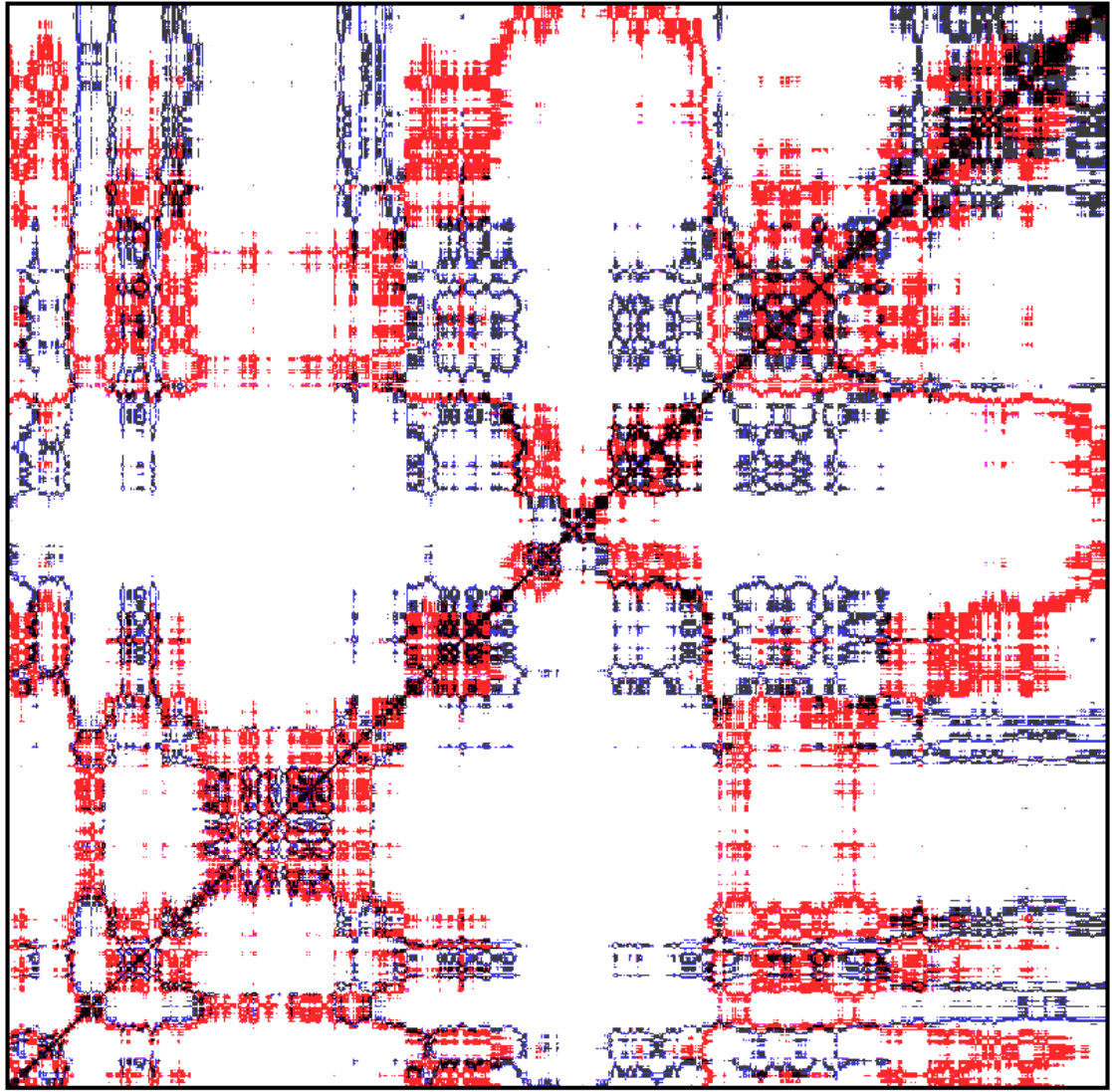

**Figure 2** - Here the two recurrence plots of Figure 1 are shown together. In blue we see the exchange rate Euro - US Dollar and in red the exchange rate Euro - GB Pound. We can see the periods of time when the two rates oscillate or are on random walk as often coincident.

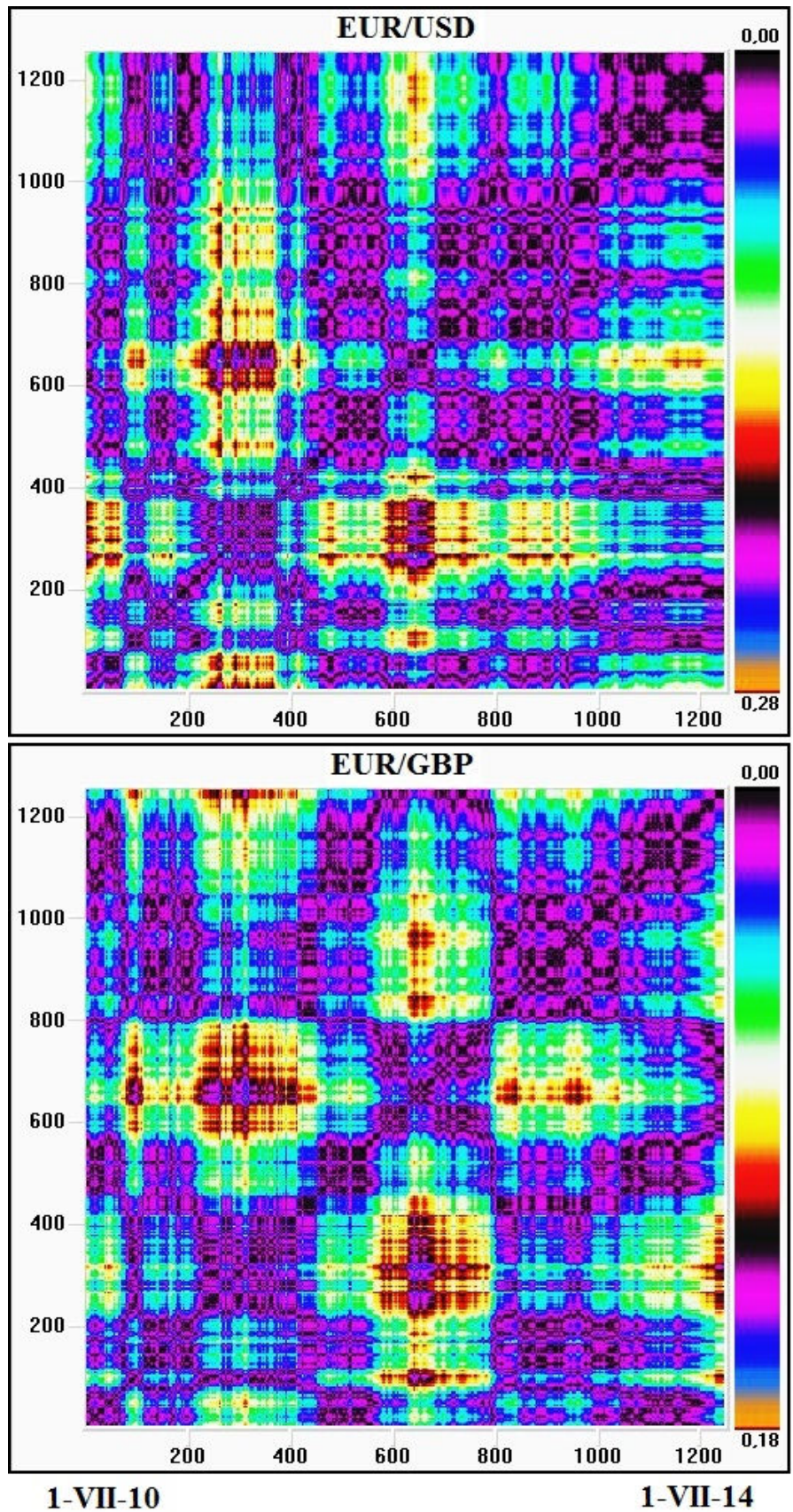

**Figure 3** - Daily exchange rate Euro - US Dollar and Euro - GB Pound in the recurrence plots created using VRA with Euclidean distances.

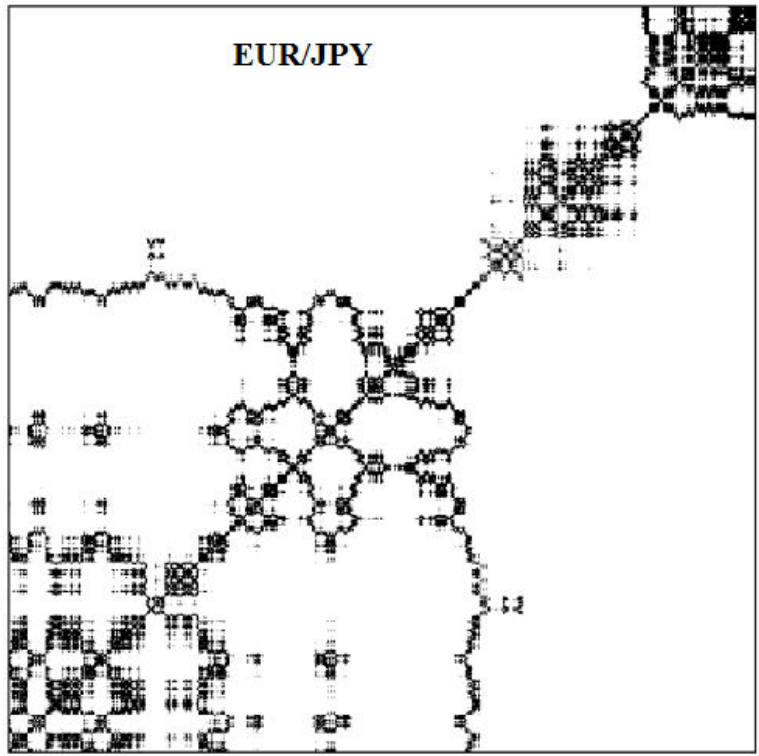
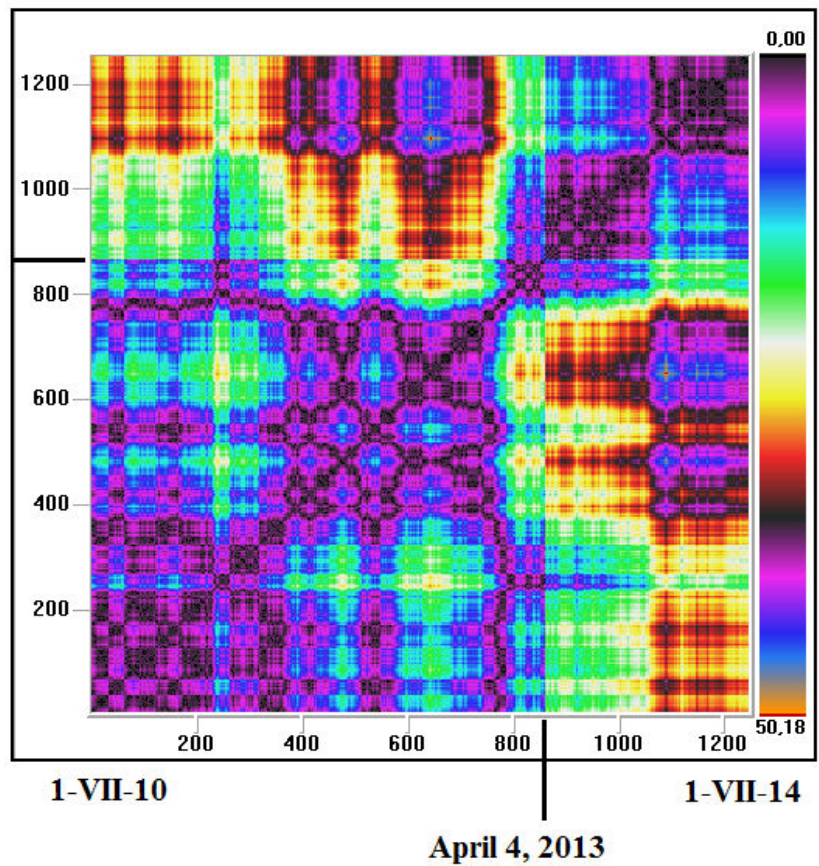

**Figure 4** - Recurrence plots of exchange rates between Euro and Japanese Yen: in the upper image, plotted according to Eqs.3 and 4, in the lower image plotted with VRA. The VRA image shows a texture transition, that is a visual discontinuity, on April 4, 2013. On that day, markets boosted by announcements of the Bank of Japan. As reported in [25], "yen fell more than 3 percent against the dollar and 4 percent against the euro".